\documentstyle[12pt]{ioplppt}

\begin{document}
\jl{3}

\date{\today}

\title{The evolution of Bernstein modes in quantum wires with
       increasing deviation from parabolic confinement}[Bernstein
       modes in quantum wires]

\author{Arne Brataas\dag\, Vidar Gudmundsson\ddag\,
        A G Mal'shukov\S\ and K A Chao\dag}
\address{\dag\ Department of Physics, Norwegian University of Science and
Technology, N-7034 Trondheim, Norway.}
\address{\ddag\ Science Institute, University of Iceland, Dunhaga 3,
         IS-107 Reykjavik, Iceland.}
\address{\S\ Institute of Spectroscopy, Russian Academy of Sciences,
         142092 Troitsk, Moscow Region, Russia}

\begin{abstract}
  We investigate the evolution of the interaction of the
  magnetoplasmon resonance with the harmonics of the cyclotron
  resonance as the confinement of an electron gas in a quantum wire
  increasingly deviates from the parabolic case. The occurrence of the
  Bernstein modes is observed in a time-dependent Hartree model of a
  two-dimensional electron gas in a single quantum wire.
\end{abstract}


\section{Introduction}

Quantum wires are interesting for probing the electron-electron
interaction in the transition regime from quasi two-dimensional
systems to quasi one-dimensional systems.  These systems can be
fabricated starting from two dimensional semiconductor microstructures
(2DES), for example in {AlGaAs-GaAs} heterostructures.  The wires may
be created by a gate structure or by etching techniques.  The presence
of a magnetic field is manifest by many interesting quantum features,
most notably the quantum Hall effect and the fractional quantum Hall
effect.  Far infrared response is a powerful method for determining
the charge-density spectra in the transverse direction.  For
transverse parabolic potentials, the motion of the electrons may be
separated into center-of-mass motion and relative motion.  The
center-of-mass motion is decoupled from the relative motion of the
electrons, and takes place in a transverse potential with the
effective frequency $(\omega_0^2+\omega_c^2)^{1/2}$, where $\hbar
\omega_0$ is the confinement energy and $\omega_c$ is the cyclotron
frequency.  The time-dependent external potential $\phi_e$
representing the long wave length far-infrared radiation only couples
to the center-of-mass motion causing a single peak in the FIR spectrum
at the plasmon frequency $\Omega_p=(\omega_c^2 + \omega_0^2)^{1/2}$
equal to the effective transverse
frequency\cite{Kohn61:1242,Pfannkuche93:6}.  Therefore, according to
Kohns theorem\cite{Kohn61:1242}, although a rich spectrum would be
expected, there is only a single resonance.  For small systems with
general confinement potentials deviating strongly from the parabolic
case, the single-particle excitations may overlap with the collective
excitations causing a fine-structure in the plasmon
peak\cite{Malshukov95:7669,Gudmundsson95:17744}. Any deviation from
parabolic confinement will couple the internal and collective motion.

Bernstein showed\cite{Bernstein58:10} that in 3D homogeneous systems
the dispersion of the magnetoplasmon as a function of the magnetic
field will have anticrossings at finite wave vectors ($\vec{k} \perp
\vec{B}$) around $\Omega_p=n \omega_c$ where $n=2,3,\cdots$.  The
calculation was based on solving the classical Boltzmann equation and
the Maxwell's equations self-consistently.  The strength of the
anticrossings depends on temperature, wave length and magnetic field
in a complex way and decreases for the higher harmonics.  An
interaction between the magnetoplasmon resonance and the harmonics of
the cyclotron resonance, a so-called Bernstein
mode\cite{Bernstein58:10}, have been observed in
3DES\cite{Patel68:1563}, 2DES\cite{Batke85:2367,Batke86:6951}, 1DES
(wires) and 0DES (dots)\cite{Gudmundsson95:17744}.  For confined
electron systems in reduced dimensions the breakdown of Kohns theorem
in nonparabolic potentials is sufficient to observe the Bernstein
modes instead of considering absorption at a finite wave vector.

We study the far infrared absorption in the integer quantum Hall
regime.  Here we consider a broad range of deviations from parabolic
wires in the time-dependent Hartree approximation.  Increasing
deviations give richer excitation spectra when the magnetoplasmon
resonance starts to couple to increasingly higher harmonics of the
cyclotron resonance.  We systematically investigate how the splitting
and its location depend on the strength of the deviation from the
parabolic confinement.

\section{Model}

We consider a strictly two-dimensional electron system lying in the
$x$-$y$ plane. The motion in the $z$ direction is neglected since the
electrons are confined to the lowest subband at the low temperature
attained in experiments.  We use the Hartree approximation (HA) to
reduce the many-particle Hamiltonian to a single-particle Hamiltonian
for each electron in an effective potential approximating the
electron-electron interaction.  For the FIR absorption we use the
corresponding time-dependent approximation describing the
self-consistent linear response of the 2DES to an external homogeneous
time-varying electrical field (the random phase approximation, RPA).
In a constant perpendicular magnetic field $B$ the cyclotron frequency
and the magnetic length are $\omega_c=eB/(m^*c)$ and
$l_c=[\hbar/(m^*\omega_c)]^{1/2}$, respectively, where $m^*$ is the
effective mass. The dielectric constant of the surrounding medium is
noted by $\kappa$.  It is convenient to introduce the constant
perpendicular magnetic field with the vector potential in the Landau
gauge $\vec{A}(\vec{r}) = (-By,0)$.

The effective Hamilton operator for a single electron in the
confining potential $V_{c}(y)$ is
\begin{equation}
      H = - \frac{\hbar^2}{2 m^*} \left( \nabla^2 - \frac{2i}{l_c^2} y
      \frac{\partial}{\partial x} \right) + \frac{1}{2} m^* \omega_c^2 y^2 +
      V_c(y) + V_{H}(\vec{r}) \, ,
\label{sinham}
\end{equation}
where $V_{H}(\vec{r})$ is the self-consistent Hartree potential
representing the direct Coulomb interaction between one electron
and the total charge density of the 2DES.
Periodic boundary condition in the longitudinal direction of the wire
gives a Bloch-type single-particle wave function
\begin{equation}
      \Psi_{nk}(x,y) = \frac{1}{\sqrt{L_x}} e^{i k x}\psi_{nk}(y) \, ,
      \label{Bloch}
\end{equation}
where the longitudinal wave vector $k = p \cdot 2 \pi/L_x$ with $p\in
Z$ and $n$ is the transverse quantum number.  The length of the wire,
$L_x$, is assumed to be much larger than the effective width of the
wire, therefore the Hartree potential only depends on the transverse
coordinate\cite{Gerhardts88:845}
\begin{equation}
      V_H(y)  = - \frac{2e^2}{\kappa} \int_{-\infty}^{\infty} dy'\: n_s(y')
      \ln{\left|\frac{y - y^{\prime}}{L} \right|} \, .
\end{equation}
with $L$ given below.  It has been assumed that a neutralising
background charge exists.  For noninteracting electrons in a simple
one-dimensional parabolic potential
$V(y)=\frac{1}{2}m^*\omega_0^2y^2$, the single-particle eigenfunctions
are
\begin{equation}
      \phi_{nk}(y) = \frac{1}{\sqrt{L}}
      \frac{1}{\sqrt{2^n n! \sqrt{\pi}}} H_n\left( \frac{y-y_k}{L}\right)
      \exp{\left[ -\frac{(y-y_k)^2}{2L^2}\right] } \, ,
\label{sinwave}
\end{equation}
where $y_k = k \cdot l_c^2 l_0^4 / (l_0^4 + l_c^4)$ is the center
coordinate, which generally does not equal the expectation value of
the transverse coordinate $y$\cite{Gudmundsson88:453}.  The $n$th
Hermite polynomial is denoted by $H_n$.  The confinement length is
defined as $l_0 = [\hbar /( m^* \omega_0)]^{1/2}$.  The electron is
localised within the effective length $L = [\hbar/(m^*\Omega)]^{1/2}$,
replacing the magnetic length $l_c$, and the cyclotron frequency
$\omega_c$ is replaced by an effective frequency defined by $\Omega =
(\omega_0^2 + \omega_c^2)^{1/2}$.  The eigenenergies corresponding to
the eigenstates (\ref{sinwave}) are $E_{nk} = \hbar \Omega (n + 1/2) +
(\hbar^2 k^2/2m^*) \times l_c^4/(l_c^4 + l_0^4) $.  In the case of
vanishing confinement, (2DES) $l_0 \rightarrow \infty$, the energy
bands reduce to the familiar dispersionless Landau levels.  The
effective single-particle Hamiltonian is diagonalised using the wave
functions of the noninteracting electrons (\ref{sinwave}) as a
functional basis and the self-consistent solutions are obtained by
iteration.

The response to a time-dependent perturbation may be found by the
random phase approximation where exchange and correlation effects are
neglected.  In this mean-field approximation the noninteracting
Hartree quasi-particles move in a self-consistent potential given by
the external perturbation and the response of the charge density of
the electrons $n_s$.  The selfconsistent potential,
$\phi_s(\vec{r},\omega)$, may be found from the external potential
$\phi_e(\vec{r},\omega)$
\begin{equation}
      \langle \alpha | \phi_s(\omega) | \beta \rangle
      =
      \langle \alpha | \phi_e(\omega) | \beta \rangle +
      \sum_{\delta \gamma}
      \frac{H_{\alpha,\beta; \gamma,\delta}(f_{\gamma}-f_{\delta})}
      {\hbar \omega + (\epsilon_{\gamma}-\epsilon_{\delta})+i0^+}
      \langle \delta | \phi_s(\omega) | \gamma \rangle \, ,
\label{selfpot}
\end{equation}
where $\alpha$ denotes the longitudinal and transverse quantum numbers
$n$ and $k$. $f_{\alpha}$ is the Fermi occupation factor and the
Hartree matrix elements are defined in terms of the Hartree ground
state wave functions
\begin{equation}
      H_{\alpha, \beta; \gamma, \delta} =
      \frac{e^2}{\kappa} \int d\vec{r} \int d\vec{r}^{\prime}
      \frac{\psi_{\gamma}^*(\vec{r}^{\prime}) \psi_{\delta}(\vec{r}^{\prime})
      \psi_{\alpha}^*(\vec{r}) \psi_{\beta}(\vec{r})}
      {|\vec{r}-\vec{r}^{\prime}|} \, .
\end{equation}
A method to find $\phi_s$ based on a Fourier transformation of the
position coordinates and repeatedly solving (\ref{selfpot}) for all
the needed values of $\omega$ is shown in Ref.\
\cite{Gudmundsson95:17744}. Here we present an alternative method
transforming (\ref{selfpot}) such that $\phi_s(\omega)$ can be
simultaneously obtained for all values of $\omega$ from a matrix
eigenvalue problem. The latter method considerably speeds up numerical
calculations and we have checked that they deliver exactly the same
results.  The equation for the selfconsistent potential
(\ref{selfpot}) simplifies for transverse external fields.  The
longitudinal quantum number $k$ is conserved in the transition, so
that each single electron transition may be labeled by $k$ and two
transverse quantum numbers $n$ and $m$.  By applying a time dependent
but spatially constant electric field, $\vec{E}_e(\vec{r},t) =
-\hat{y}{\cal E}_e \exp (-i \omega t)$, the Fourier component of the
external potential is $\phi_e(\vec{r},\omega) = -e {\cal E}_{e} y$,
where $-e$ is the electron charge.  In the basis (\ref{Bloch}) the
interband matrix elements of the external potential $\{\phi_e(\omega)
\}_{n,m}^k=\int dy \psi_{nk}(y) \phi_e(\vec{r},\omega) \psi_{mk}(y)$
are real and symmetric.  Since the Hartree potential is local the
matrix elements of the self-consistent potential are also symmetric
with respect to the transverse quantum numbers
$\{\phi_s(\omega)\}_{n,m}^{k} = \{ \phi_s(\omega)\}_{m,n}^{k}$.  This
is also the case for local-density approximations, but not for the
Hartree-Fock approximation, since the Fock potential is
nonlocal\cite{Brataas95:xx}.  The equation for the self-consistent
potential may then be written as an eigenvalue problem
\begin{eqnarray}
      & &\left(\hbar^2 \omega^2 - (\epsilon_{n,m}^k)^2 \right)
      \left\{ \eta(\omega) \right\}_{n,m}^k
      =
      \sqrt{2f_{n,m}^k \epsilon_{n,m}^k}
      \left\{\phi_e(\omega)\right\}_{n,m}^k \nonumber \\
      & - &
      2 \sum_{n^{\prime},m^{\prime},k^{\prime}}
      ^{\epsilon_{n^{\prime},m^{\prime}}^{k^{\prime}} > 0}
      H_{n,m;n^{\prime},m^{\prime}}^{k;k^{\prime}}
      \sqrt{f_{n,m}^k \epsilon_{n,m}^k}
      \sqrt{f_{n^{\prime},m^{\prime}}^k
      \epsilon_{n^{\prime},m^{\prime}}^{k^{\prime}}}
      \left\{ \eta(\omega) \right\}_{n^{\prime},m^{\prime}}^{k^{\prime}} \, ,
      \label{RPA}
\end{eqnarray}
where $\epsilon_{n,m}^k = \epsilon_n^k-\epsilon_m^k$ are the
quasi-particle excitation energies and $f_{n,m}^k = f_m^k-f_n^k$ are
the differences in occupation between initial and final states.  For
convenience we have introduced the variable $\{\eta(\omega)\}_{n,m}^k
= \{ \phi_s(\omega) \}_{n,m}^k (2f_{n,m}^k
\epsilon_{n,m}^k)^{1/2}/(\hbar^2 \omega^2-(\epsilon_{n,m}^k)^2)$.  The
resonances in the FIR spectrum are the eigenvalues of the symmetric
matrix
\begin{equation}
A_{n,m;n^{\prime},m^{\prime}}^{k;k^{\prime}} =
(\epsilon_{n,m}^k)^2
\delta_{n,m;n^{\prime},m^{\prime}}^{k;k^{\prime}} - 2
H_{n,m;n^{\prime},m^{\prime}}^{k;k^{\prime}} (f_{n,m}^k
\epsilon_{n,m}^k f_{n^{\prime},m^{\prime}}^{k^{\prime}}
\epsilon_{n^{\prime},m^{\prime}}^{k^{\prime}})^{1/2} \, .
\end{equation}
The power absorption may be found from the Joule heating
\begin{equation}
      P(\omega) = \frac{1}{2} \int d\vec{r} \: \mbox{Re} \left\{ \delta
      \vec{J}(\vec{r},\omega) \vec{E}_s^*(\vec{r},\omega) \right\} \, ,
\end{equation}
where $\delta \vec{J}(\vec{r},\omega)$ is the induced current in the
wire and $\vec{E}_s$ is the self-consistent electric field.  It is
\begin{eqnarray}
      P(\omega) & = &
      - \frac{\omega}{2} \sum_{n,m,k}^{\epsilon_{n,m}^k > 0}
      \left \{ \phi_e(\omega) \right\}_{n,m}^k \mbox{Im}
      \left[
      \frac{2 f_{n,m}^k \epsilon_{n,m}^k}
      {(\hbar \omega)^2 - (\epsilon_{n,m}^k)^2}
      \left\{ \phi_s(\omega)
      \right\}_{n,m}^k
      \right] \nonumber \\
      & = &
      - \frac{\omega}{2} \sum_{n,m,k}^{\epsilon_{n,m}^k > 0}
      \sqrt{2 f_{n,m}^k \epsilon_{n,m}^k}
      \{ \phi_e(\omega) \}_{n,m}^k \mbox{Im}
      \{ \eta(\omega)   \}_{n,m}^k \, .
\label{FIR}
\end{eqnarray}
The Coulomb interaction may shift the poles of the matrix
$\{\eta(\omega)\}_{n,m}^k$ from the Hartree single-particle
excitations $\epsilon_{n,m}^k$ as can be seen from the eigenequation
(\ref{RPA}).  The new poles will show up as resonances in the FIR
absorption (\ref{FIR}).  Here a phenomenological broadening should be
inserted to give the lifetime of the excitations.  In a Fermi liquid
the scattering cross-section decreases for lower energy one-particle
excitations.  Hence, one should expect that at least for low-energy
one-particle excitations this broadening is small.  The power
absorption may be expanded in the eigenfunctions of the matrix $A$.
The oscillator strengths can then be found and the whole FIR spectrum
determined.  Since the Hartree-RPA is a conserving
approximation\cite{Baym61:287}, the longitudinal f-sum rule for the
oscillator strengths is satisfied for arbitrary electron-electron
interaction strengths.

We study quantum wires
with confining potentials of the form
\begin{equation}
  V_c(y) = \frac{1}{2} \hbar \omega_0 \left[ \left(\frac{y}{l_0}
    \right)^2 + a \left(\frac{y}{l_0}\right)^4 + b
    \left(\frac{y}{l_0}\right)^6 \right] \, ,
\end{equation}
where $a$ and $b$ are parameters that determine the higher order
deviations from parabolic confinement.  By the method described above,
we are able to find the FIR spectra arising from this confining
potential.

\section{Numerical results}

The functional basis of the Hartree ground state has been chosen large
enough so that a further expansion or iteration of the
Hartree-equations does not result in visual changes to the
single-particle energy spectra or the electron density $n_s(\vec r)$
of the ground state.  To attain sufficient accuracy in the calculation
of the absorption the size of the functional basis for the excited
states has been chosen such that further refinement results in changes
to the location of the absorption peaks that are smaller than a
typical linewidth in experiments 0.1 meV.

For the calculations we employ the usual GaAs parameters,
$m^*=0.067m_0$, $\kappa=12.4$, where $m_0$ is the the free electron
mass.  The absorption is only weakly dependent on $T$ in the
temperature range $T\leq 4\: $K for the parameters we use.  The
calculations have thus been performed for $T=1.0\: $K.

The FIR spectra have been calculated in the range of magnetic fields
$B=0-3\:$T.  For a strict parabolic confinement, $a=0$ and $b=0$, we
have checked that the generalised Kohns theorem is satisfied with a
high degree of accuracy.  We consider pure fourth order deviations,
$a=0.01$ to $a=0.40$, and pure sixth order deviations, $b= 0.001$ to
$b=0.030$.  Mixed deviations of fourth and sixth order do not give
significant new qualitative information.

For small deviations, with either $a\neq 0$ or $b\neq 0$, a single
anticrossing close to the $2\omega_c$ line appears.  The position of
the anticrossing as a function of the deviation $a$ or $b$ is shown in
Fig.\ \ref{abdev}.  The horizontal line shows where
$\Omega_p=2\omega_c$, i.e. where a possible resonance between the
magnetoplasmon and the cyclotron frequency should appear.  The
calculated anticrossing is to the right of the $2\omega_c$-line for
small deviations, but is shifted quite strongly to the left with
increasing deviations.  It is due to the interaction of the collective
oscillations, the magnetoplasmons, with the first harmonic of the
cyclotron resonance, $2\omega_c$, a so called Bernstein
mode\cite{Bernstein58:10}.

The absorption peaks as a function of the magnetic field for
increasing fourth order deviation, $a$=0.03, 0.10, 0.20, 0.40, are
shown in Fig.\ \ref{adev} and for increasing sixth order deviation,
$b$=0.003, 0.005, 0.010, 0.030, in Fig.\ \ref{bdev}.  Only absorption
peaks with an oscillator strength more than 1\% of the total
oscillator strength are shown.  The absorption peaks are labeled
strong, weak, very weak and extremely weak according to their relative
strength for a given magnetic field.  Only excitation energies in the
range 3.5-8.5 meV are shown.  The location of the strong resonance at
zero magnetic field increases in energy with increasing deviation,
because of the stronger confinement.  However for intermediate
magnetic fields the strong resonance develops a slightly negative
slope with respect to the magnetic field.  This softening of the mode
increases with increasing deviations.  For large deviations the simple
splitting develops into a complex one, see Fig.\ \ref{adev} for a
fourth order deviation and Fig.\ \ref{bdev} for a sixth order
deviation.  We see that strong deviations from a parabolic confinement
leads to interactions with higher cyclotron harmonics, $n \omega_c$,
where $n$ is larger than 2.  The new modes may also interact with each
other.  However it is not possible from the spectra to say which
harmonics are involved, since several harmonics can be excited in a
complicated way, and the anticrossing occurs at or in between the
$n\omega_c$ lines.  The spectrum is now very rich, because of the
strong coupling of the collective motion to the internal motion in the
wire.  Splittings may also be seen on the weak resonance branches for
$a=0.40$ in Fig.\ \ref{adev} and for $b=0.030$ in Fig.\ \ref{bdev}.

The intensity as a function of the magnetic field and the supplied
energy from the external potential is showed in Fig.\ \ref{a020} for a
pure fourth order deviation, $a=0.20$.  Only resonances that have a
strength of more than 1\% are shown.  We see that for this deviation
there is mainly one anticrossing, all other resonances are weak.  The
softening of the modes occurs as long as two main resonances are seen,
i.e as long as the modes are coupled.

The effects of the deviation is larger for a higher density of the
electrons, since the chemical potential is higher and the
2DEG occupies states farther from the center of the wire that are
more affected by the fourth and
sixth order terms of the confinement.

\section{Summary}

The FIR spectrum for a wire with a confining potential deviating from a
parabolic one was calculated in the time-dependent Hartree approximation.
The anticrossing due to interaction of the magnetoplasmon with
the first harmonic of the cyclotron resonance
shifts strongly from the right of the
$2\omega_c$ intersection with the plasmon dispersion to the left with
increasing deviations and the
simple splitting develops into a more complex one, including interactions
with higher harmonics of the cyclotron resonance.

In the quantum wire $\hbar\omega_c$ is not a characteristic single
electron excitation energy, but the width of the system defines a
characteristic wave vector which does not vanish with a decreasing
deviation from the parabolic confinement. An external perturbation
does excite plasma waves with a broad range of wave vectors, but the
system responds in the strongest fashion for integer multiples of the
'natural wave vector'. In a homogeneous two-dimensional electron
system the response is always with the same frequency and wave vector
as the external perturbation, thus the asymptotic behaviour of the
split-off modes is simple.  Here the asymptotic behaviour is
complicated by the fact that the split-off modes again interact with
higher order plasmons.  A detailed study of this phenomena will be
published elsewhere.

\ack{This research was supported in part by the Icelandic
Natural Science Foundation, the University of Iceland Research Fund,
and a NorFA Network Grant.}

\Figures

\begin{figure}
\caption{The location of the first anticrossing
  as a function of the magnetic field $B$ and the fourth order
  deviation $a\neq 0$, $b=0$ ($\Diamond$) or the sixth order deviation
  $a=0$, $b\neq 0$ ($+$).  The horizontal line shows where
  $\Omega_p=2\omega_c$.  $30$ states are occupied, $L_x=240\: $nm,
  $\hbar\omega_0=3.94\: $meV, $T=1.0\: $K, $m^*=0.067m_0$, $\kappa =
  12.4$.}
\label{abdev}
\end{figure}

\begin{figure}
\caption{The location of the absorption peaks
  as a function of the magnetic field $B$ and the energy
  $E=\hbar\omega$ supplied by $\phi_e$ for a pure fourth order
  deviation.  Only absorption peaks with an oscillator strength larger
  than 1\% of the total oscillator strength are shown.  The straight
  lines represent the harmonics of the cyclotron frequency, $n \times
  \omega_c$ ($n$=2,3,4,5).  Strong absorption is marked by $\Diamond$,
  weak one by $+$, very weak one by $\Box$, and extremely weak one by
  $\times$.  $30$ states are occupied, $L_x=240\: $nm,
  $\hbar\omega_0=3.94\: $meV, $T=1.0\: $K, $m^*=0.067m_0$, $\kappa =
  12.4$.}
\label{adev}
\end{figure}

\begin{figure}
\caption{The location of the absorption peaks as a function of the
  magnetic field $B$ and the energy $E=\hbar\omega$ supplied by
  $\phi_e$ for a pure sixth order deviation.  Only
  absorption peaks with an oscillator strength larger than 1 \% of the
  total oscillator strength are shown.  The lines are the possible
  interaction of the magnetoplasmon with harmonics of the cyclotron
  frequency, $n \times \omega_c$ ($n$=2,3).  Strong absorption is
  marked by $\Diamond$, weak one by $+$, and very weak one by $\Box$.
  $30$ states are occupied, $L_x=240\: $nm, $\hbar\omega_0=3.94\:
  $meV, $T=1.0\: $K, $m^*=0.067m_0$, $\kappa = 12.4$.}
\label{bdev}
\end{figure}

\begin{figure}
\caption{The intensity of the absorption peaks as a function of the
  magnetic field $B$ and the energy $E=\hbar\omega$ supplied by
  $\phi_e$ for a pure fourth order deviation $a=0.20$.
  Only absorption peaks with an oscillator strength larger than 1 \%
  of the total oscillator strength are shown.  $30$ states are
  occupied, $L_x=240\: $nm, $\hbar\omega_0=3.94\: $meV, $T=1.0\: $K,
  $m^*=0.067m_0$, $\kappa = 12.4$.}
\label{a020}
\end{figure}

\end{document}